\documentclass[12pt]{article}

\newcommand{\bea}{\begin{eqnarray}}
\newcommand{\eea}{\end{eqnarray}}
\newcommand{\be}{\begin{equation}}
\newcommand{\ee}{\end{equation}}
\newcommand{\vs}[1]{\vspace{#1 mm}}

\newcommand{\dsl}{\pa \kern-0.5em /}

\newcommand{\pa}{\partial}

\newcommand{\nn}{\nonumber\\}

\newcommand{\eqn}[1]{(\ref{#1})}

\newcommand{\G}{{\bar G}}

\newcommand{\thet}{{\tilde \theta}}
\newcommand{\rh}{{\tilde \rho}}

\begin{document}
\topmargin 0pt
\oddsidemargin 0mm

\begin{flushright}

USTC-ICTS-09-04\\




\end{flushright}

\vspace{2mm}

\begin{center}

{\Large \bf Non-supersymmetric D1/D5, F/NS5
and closed string tachyon condensation}
\vs{6}

{\large J. X. Lu$^a$\footnote{E-mail: jxlu@ustc.edu.cn}, Shibaji
Roy$^b$\footnote{E-mail: shibaji.roy@saha.ac.in}, Zhao-Long
Wang$^a$\footnote{E-mail: zlwang4@mail.ustc.edu.cn} and Rong-Jun
Wu$^a$\footnote{E-mail: rjwu@mail.ustc.edu.cn} }

 \vspace{4mm}

{\em

 $^a$ Interdisciplinary Center for Theoretical Study\\

 University of Science and Technology of China, Hefei, Anhui
 230026, China\\




\vs{4}

 $^b$ Saha Institute of Nuclear Physics,
 1/AF Bidhannagar, Calcutta-700 064, India}

\end{center}


\begin{abstract}

We construct the intersecting non-supersymmetric (non-susy) D1/D5
solution of type IIB string theory. While, as usual, the solution is
charged under an electric two-form and an electric six-form gauge
field, it also contains a non-susy chargeless (non-BPS) D0-brane.
The S-dual of this solution is the non-susy F/NS5 solution. We show
how these solutions nicely interpolate between the corresponding
black (or non-extremal) solutions and the Kaluza-Klein (KK) ``bubble
of nothing'' (BON) by continuously changing some parameters
characterizing the solutions from one set of values to another. We
show, by a time symmetric general bubble initial data analysis, that
the final bubbles in these cases are static and stable and the
interpolations can be physically interpreted as closed string
tachyon condensation. As  special cases, we recover the transition
of two charge black F-string to BON, considered by Horowitz, and
also the transition from AdS$_3$ black hole to global AdS$_3$.

\end{abstract}

\newpage

\section{Introduction}

In an interesting paper it has been argued by Horowitz
\cite{Horowitz:2005vp} that under certain conditions black strings
of type II string theories have dramatic new endpoints to Hawking
evaporation in the form of KK BON \cite{Witten:1981gj}\footnote{This
has also been generalized to the $p$ = 0 case, i.e., charged or
uncharged black hole case, in \cite{Green:2006nv}.}. He arrived at
this conclusion by applying the closed string tachyonic instability
to black strings. Closed string tachyons are known to develop when
the fermions in the theory are taken to satisfy antiperiodic
boundary conditions along one of the compact directions and the size
of the circle becomes of the order of string scale
\cite{Scherk:1978ta,Rohm:1983aq}. Adams et. al. \cite{Adams:2005rb}
have argued that when these winding string tachyons are localized
they can trigger a topology changing transformation as a consequence
of the closed string tachyon condensation. The transition from the
black string to the KK BON is an application of this process. A
similar transition also occurs for black D$p$-branes and was briefly
mentioned in \cite{Horowitz:2005vp}. However, it was observed that
only for $p = 3$ the final bubble could be static and stable. In
ref.\cite{Lu:2007bu} we showed that this is not quite right and in
fact, black D$p$-branes can make transitions to stable static
bubbles for all $p \leq 4$ via closed string tachyon condensation
but in an indirect way as specified there for $p \neq 3$. We have
also argued how this stringy process can be modelled by a series of
classical supergravity configurations.

In \cite{Horowitz:2005vp} Horowitz started from the standard
black fundamental string solution \cite{Horowitz:1991cd,Duff:1993ye}
and compactified the string direction. Since here the metric
is multiplied appropriately with a harmonic function, the size of
the string wound along the compact direction varies monotonically
from $L$ (where $L$ is the periodicity of the compact direction) to
zero as we move along the radial direction from infinity to the
singular point. So, at some point in between the size of the circle
becomes of the order of string scale and if this occurs on the
horizon, then the closed string tachyon condensation causes the
circle to pinch off and the resulting state is a bubble which in
this case cannot be static but should expand out. Static bubble
appeared while considering a similar transition for a toroidally
compactified black F-string solution containing both F and NS5 charge.
However, in order to show that bubble is the end state of this
transition, a time symmetric bubble initial data analysis has been
performed and it was found that indeed under certain
conditions ($Q/L^2\ll 1$, where $Q$ is the flux associated with the
bubble) the bubble can be stable and so, the black F-string in those
cases can make a transition to classically stable static bubbles. In
this analysis the two charges of the black F-string were made equal,
for simplicity, which in turn, decouples the dilaton in the solution.

In this paper we will construct the non-susy F/NS5 solution of the
ten dimensional type IIB string theory with two unequal charges and
a non-trivial dilaton, generalizing the solution considered by
Horowitz. In order to construct this solution we will start from the
known type IIA solution representing intersections of charged
non-susy\footnote{Note that unlike the BPS branes the non-susy
branes could be either charged or chargeless.} D0-brane and charged
non-susy D4-branes given in \cite{Bai:2005jr}. We next obtain a
delocalized (in one of the transverse directions of D4-brane)
version of this solution which will introduce an isometry direction
and then take T-duality along that direction. Note that since we are
dealing with non-susy solution the procedure of delocalization and
the resulting T-dual configurations are quite different from the
usual BPS solutions we are familiar with. In this case (as opposed
to BPS case) after T-duality, we obtain the intersecting charged
non-susy D1/D5 solution and chargeless non-susy or non-BPS
D0-branes. The S-dual of this solution is the intersecting non-susy
F/NS5 solution. This solution can be interpreted as intersecting
charged non-susy F/NS5 solution along with chargeless non-susy or
non-BPS D0-branes. Both these solutions are characterized by six
independent parameters. We will see that when these parameters
are varied from one set of values to another keeping the physical
conserved quantities such as the mass and the charges unchanged,
these solutions
nicely interpolate between black solutions and KK BON.

We emphasize that the existence of interpolating classical solutions does not
necessarily imply that the transition from black brane to KK BON will
actually occur (as happened for D5- and D6-branes).
As we mentioned, in order to have a transition, we must ensure that the
final bubble configuration is locally stable so that it does not evolve
further perturbatively and this is done by the time symmetric general bubble
initial data analysis. To show that the transition is caused by a
perturbative
process such as the closed string tachyon condensation, further conditions
have to be satisfied. In particular, the curvature of the black
brane
near the horizon (where the closed string tachyon condensation occurs)
must be much smaller than the string scale, otherwise the classical
description breaks down and the black brane makes a
transition to open string modes. Also, the horizon size and the bubble size,
the charge of the black brane and the flux of the bubble and
finally,
the size of the compact circle at infinity of both the configurations must
be equal. If all these requirements are satisfied, we can conclude that
the black
brane can make a transition to KK BON through closed string tachyon
condensation. Conversely, if a black configuration makes a
transition to KK BON via closed string tachyon condensation with all the
restrictions being satisfied, the classical interpolating solution, if it
exists, can be regarded as a model for the stringy process like the closed
string tachyon condensation.

This is precisely what we will show in this paper. After constructing the
interpolating solutions we will perform a time symmetric general bubble
initial data analysis with unequal charges and non-trivial dilaton (as opposed
to the case considered by Horowitz) to show that the final bubble
configuration can indeed be classically stable. This in turn implies that
the interpolation means really a transition from black-brane to bubble. Then
we carefully look at the various conditions mentioned earlier for the
interpretation of this transition and whether it can be caused by a
stringy process of closed string tachyon
condensation and we find that indeed in certain cases this is true. The
interpolating classical supergravity solutions can then be regarded as a model
for this process. As a special case of non-susy F/NS5, the black-string to
KK BON transition for the two charge black F-string in $D=6$ considered by
Horowitz can be understood and also from non-susy D1/D5, the AdS$_3$ black
hole to AdS$_3$ soliton transition can be understood as a special case.

\section{The interpolating solutions}

In this section we will construct both the non-susy D1/D5 and F/NS5 solutions
and show how by varying a subset of the parameters characterizing
the solutions
they nicely interpolate between the corresponding black or non-extremal
solutions and the KK BON. For this purpose we start from
the intersecting charged non-susy D$(p-4)$/D$p$ solution given in
eqs.(1) -- (6) of ref.\cite{Bai:2005jr} for $p=4$. The corresponding
charged non-susy D0/D4 solution is given as,
\bea\label{D0D4}
ds^2 &=& F_2^{-\frac{3}{8}} F_1^{-\frac{7}{8}}(-dt^2) +
  F_2^{-\frac{3}{8}} F_1^{\frac{1}{8}}\sum_{i=1}^4 (dx^i)^2
+ \left(H\tilde{H}\right)^{\frac{2}{3}} F_2^{\frac{5}{8}}
F_1^{\frac{1}{8}}\left(dr^2 + r^2 d\Omega_4^2\right),\nn e^{2\phi}
&=& F_2^{-\frac{1}{2}} F_1^{\frac{3}{2}} \left(\frac{H}{\tilde
    H}\right)^{2\delta_1},\nn
F_{[4]} &=& b {\rm Vol}(\Omega_4), \qquad F_{[8]} = c {\rm
Vol}(\Omega_4) \wedge dx^1 \wedge dx^2 \wedge dx^3 \wedge dx^4, \eea
where \be\label{functions} F_{1,2} = \cosh^2\theta_{1,2}
\left(\frac{H}{\tilde H}\right)^{\alpha_{1,2}} - \sinh^2\theta_{1,2}
\left(\frac{\tilde H}{H}\right)^{\beta_{1,2}} \ee with
$H=1+\omega^3/r^3$ and $\tilde H = 1 - \omega^3/r^3$, where
$r=\sqrt{ (x^5)^2 + \ldots + (x^9)^2}$. Note that the metric in
\eqn{D0D4} is given in the Einstein frame. (The metric in general
would be given in the Einstein frame unless mentioned explicitly.)
The solution is well-defined in the region $r>\omega$ and there is a
singularity at $r=\omega$. Here $\alpha_{1,2}$, $\beta_{1,2}$,
$\theta_{1,2}$, $\delta_1$, $\omega$ are integration constants and
$b$, $c$ are the charge parameters, but not all the constants are
independent. There are five relations among them given as follows,
\bea\label{parameters} & &\alpha_1 - \beta_1 = -\frac{3}{2}
\delta_1, \qquad\qquad \alpha_2 - \beta_2 = \frac{1}{2} \delta_1,\nn
& & b = 3(\alpha_2 + \beta_2) \omega^3 \sinh2\theta_2, \qquad c =
3(\alpha_1 + \beta_1) \omega^3 \sinh2\theta_1,\nn & & (\alpha_1 +
\beta_1)^2 + (\alpha_2 + \beta_2)^2 + \frac{3}{2} \delta_1^2 =
\frac{32}{3}. \eea By eliminating $\beta_{1,2}$ in the last relation
of \eqn{parameters} we can rewrite it as, \be\label{newrelation}
\frac{1}{2}\delta_1^2 + \frac{1}{2}\alpha_1(\alpha_1 + \frac{3}{2}
\delta_1) + \frac{1}{2} \alpha_2(\alpha_2 - \frac{1}{2} \delta_1) =
\frac{4}{3}. \ee So, the number of independent parameters
characterizing the solution is five and we interpreted these
parameters in \cite{Bai:2005jr} as the no. of D4-branes, no. of anti
D4-branes, no. of D0-branes, no. of anti D0-branes and the tachyon
parameter. Also note that in the solution \eqn{D0D4} both the
non-susy D4-branes and the non-susy D0-branes are magnetic. The
corresponding electric solution will have the same form as
\eqn{D0D4} with the field-strengths $F_{[8]}$ and $F_{[4]}$ replaced
by the electric gauge fields \bea A_{[1]} &=&
\frac{1}{2}\sinh2\theta_1 \left(\frac{C_1}{F_1}\right) dt,\nn
A_{[5]} &=& \frac{1}{2}\sinh2\theta_2 \left(\frac{C_2}{F_2}\right)
dt \wedge dx^1 \wedge dx^2 \wedge dx^3 \wedge dx^4 \eea where
$C_{1,2} = (H/\tilde H)^{\alpha_{1,2}} - (\tilde
H/H)^{\beta_{1,2}}$. Now if we want to have non-susy D1/D5 solution
from here we first have to create an isometry direction along which
we can take a T-duality transformation. For BPS branes this is
usually done by placing the BPS branes in a periodic array along one
of the transverse directions of the brane and then taking a
continuum limit. This is possible due to the no-force condition of
the BPS branes. For non-susy branes this procedure does not work and
we have to obtain the delocalized solution directly by solving the
equations of motion with a suitable ansatz. This was first done in
\cite{Lu:2004xi} and then later in \cite{Lu:2005ju,Lu:2005jc}. So,
from our experience we can write down the non-susy charged
intersecting D0/D4 solution delocalized in one transverse direction
as, \bea\label{delocalizedD0D4} ds^2 &=& F_2^{-\frac{3}{8}}
F_1^{-\frac{7}{8}}(-dt^2) +
  F_2^{-\frac{3}{8}} F_1^{\frac{1}{8}}\sum_{i=1}^4 (dx^i)^2 +
  F_2^{\frac{5}{8}} F_1^{\frac{1}{8}}
\left(\frac{H}{\tilde H}\right)^{2\delta_2} (dx^5)^2\nn
& & \qquad\qquad + \left(H\tilde{H}\right) \left(\frac{H}{\tilde
H}\right)^{-\delta_2} F_2^{\frac{5}{8}} F_1^{\frac{1}{8}}\left(dr^2
+ r^2 d\Omega_3^2\right),\nn e^{2\phi} &=& F_2^{-\frac{1}{2}}
F_1^{\frac{3}{2}} \left(\frac{H}{\tilde
    H}\right)^{2\delta_1},\nn
A_{[1]} &=& \frac{1}{2}\sinh2\theta_1 \left(\frac{C_1}{F_1}\right)
dt,\nn A_{[5]} &=& \frac{1}{2}\sinh2\theta_2
\left(\frac{C_2}{F_2}\right) dt \wedge dx^1 \wedge dx^2 \wedge dx^3
\wedge dx^4, \eea where $F_{1,2}$ remains the same as in
\eqn{functions}, but $H$ and $\tilde H$ have the forms $H = 1 +
\omega^2/r^2$, $\tilde H = 1 -\omega^2/r^2$. Here $r = \sqrt{(x^6)^2
+ \ldots + (x^9)^2}$. The parameter relations also remain the
same\footnote{For the magnetic solutions the parameter relations
$b$, $c$ will change as $b = 2(\alpha_2 + \beta_2) \omega^2
\sinh2\theta_2$ and $c= 2(\alpha_1+\beta_1) \omega^2
\sinh2\theta_1$.} as in \eqn{parameters} except the last one (see
the form in \eqn{newrelation}) which takes the form,
\be\label{newparameter} \frac{1}{2} \delta_1^2 + \frac{1}{2}
\alpha_1(\alpha_1 + \frac{3}{2} \delta_1) + \frac{1}{2} \alpha_2
(\alpha_2 - \frac{1}{2}\delta_1) = (1-\delta_2^2)\frac{3}{2}. \ee
The solution \eqn{delocalizedD0D4} represents intersecting non-susy
D0/D4 solution delocalized in $x^5$ direction. So, $x^5$ is an
isometry direction along which we will take a T-duality
transformation to obtain the localized intersecting non-susy D1/D5
solution. Using the standard rules of T-duality transformation
\cite{Bergshoeff:1995as, Das:1996je, Breckenridge:1996tt} on
\eqn{delocalizedD0D4} we obtain, \bea\label{D1D5D0} ds^2 &=& {\hat
F}_5^{-\frac{1}{4}} {\hat F}_1^{-\frac{3}{4}} \left(\frac{H}{\tilde
H}\right)^{\frac{\delta_1}{4} + \frac{\delta_2}{2}} (-dt^2) +
  {\hat F}_5^{-\frac{1}{4}} {\hat F}_1^{\frac{1}{4}}
\left(\frac{H}{\tilde H}\right)^{\frac{\delta_1}{4} + \frac{\delta_2}{2}}
\sum_{i=1}^4 (dx^i)^2 \nn
& & +
  {\hat F}_5^{-\frac{1}{4}} {\hat F}_1^{-\frac{3}{4}}
\left(\frac{H}{\tilde H}\right)^{-\frac{3\delta_1}{4} -
\frac{3\delta_2}{2}} (dx^5)^2 + \left(H\tilde{H}\right)
\left(\frac{H}{\tilde H}\right)^{-\frac{\delta_1}{4} -
\frac{\delta_2}{2}} {\hat F}_5^{\frac{3}{4}} {\hat
F}_1^{\frac{1}{4}}\left(dr^2 + r^2 d\Omega_3^2\right),\nn e^{2\phi}
&=& {\hat F}_5^{-1} {\hat F}_1
 \left(\frac{H}{\tilde H}\right)^{2\delta_1 - 2\delta_2},\nn
A_{[2]} &=& \frac{1}{2}\sinh2\theta_1 \left(\frac{{\hat C}_1} {{\hat
F}_1}\right) dt \wedge dx^5,\nn A_{[6]} &=&
\frac{1}{2}\sinh2\theta_5 \left(\frac{{\hat C}_5} {{\hat
F}_5}\right) dt \wedge dx^1 \wedge dx^2 \wedge dx^3 \wedge dx^4
\wedge dx^5, \eea where we have replaced the subscript `2' in the
functions ${\hat F}$ by `5' to indicate that it is associated with
the D5-brane. We have defined \be\label{newfunctions} {\hat F}_{1,5}
= \cosh^2\theta_{1,5} \left(\frac{H}{\tilde
H}\right)^{\hat{\alpha}_{1,5}} - \sinh^2\theta_{1,5}
\left(\frac{\tilde H}{H}\right)^{\hat{\beta}_{1,5}} \ee with ${\hat
C}_{1,5} = (H/\tilde H)^{{\hat \alpha}_{1,5}} - (\tilde H/H)^{{\hat
\beta}_{1,5}}$, and $H = 1 + \omega^2/r^2$, $\tilde H = 1 -
\omega^2/r^2$. The functions ${\hat F}_{1,5}$ are related to the
previous functions $F_{1,2}$ as follows, \be\label{functionrelation}
{\hat F}_1 =  F_1, \qquad {\hat F}_5 = \left(\frac{H}{\tilde
H}\right)^{\frac{1}{2}\delta_1} F_2. \ee The parameter relations are
now given as, \bea\label{parametersnew} && {\hat \alpha}_1 - {\hat
\beta}_1 = -\frac{3}{2} \delta_1, \qquad
 {\hat \alpha}_5 - {\hat \beta}_5 = \frac{3}{2} \delta_1,\nn
&& ({\hat \alpha}_1 + {\hat \beta}_1)^2 + ({\hat \alpha}_5 + {\hat
\beta}_5)^2 + \frac{3}{2} \delta_1^2 = (1-\delta_2^2)12. \eea where
the relations between the old parameters and the `hatted' new
parameters are given as, \be\label{oldnewparameter} {\hat \alpha}_5
= \alpha_2 + \frac{1}{2} \delta_1, \quad {\hat \alpha}_1 = \alpha_1,
\quad {\hat \beta}_5 = \beta_2 - \frac{1}{2}\delta_1, \quad {\hat
\beta}_1 = \beta_1. \ee By inspection it is clear from \eqn{D1D5D0},
that since the coefficient of $(-dt)^2$ is different from that of
$\sum_{i=1}^4 (dx^i)^2$, the solution contains chargeless D0-brane.
Also since $A_{[2]}$ and $A_{[6]}$ are non-zero the solution
contains charged non-susy D-string lying along $x^5$ as well as
charged non-susy D5-branes lying along $x^1, \ldots, x^5$ directions
and so, \eqn{D1D5D0} represents intersecting charged non-susy D1/D5
system with chargeless non-susy D0-brane. The solution \eqn{D1D5D0}
has six independent parameters, for example, one such independent
set is $\omega$, $\theta_{1,5}$, $(\hat{\alpha}_{1,5} +
\hat{\beta}_{1,5})$, and $\delta_2$.

To verify the correctness of the solution \eqn{D1D5D0}, we can check some
special cases. For example, if we put
$\delta_1=-2\delta_2$, then \eqn{D1D5D0} reduces to intersecting non-susy
D1/D5 brane system obtained in \cite{Bai:2005jr}. Also redefining the
paremeters as,
\be\label{redefine}
{\hat \alpha}_5 = \frac{3}{2} {\bar \delta}_1 - 2 \delta_0 + {\bar \alpha},
\quad {\hat \alpha}_1 = - {\bar \delta}_1 - 2 \delta_0,\quad \delta_1 =
{\bar \delta}_1 - \frac{8}{3} \delta_0, \quad \delta_2 = {\bar \delta}_2 +
\frac{4}{3} \delta_0
\ee
and putting $\theta_1 = 0$ along with the redefinition ${\hat F}_5 = {\bar
  F}_5 \left(H/\tilde H\right)^{\frac{3}{2} {\bar \delta}_1 - 2\delta_0}$, the
above solution \eqn{D1D5D0} reduces to charged non-susy D5 brane intersecting
with chargeless D1 and D0 branes considered in \cite{Lu:2007bu} with the
new function
${\bar F}_5 = (H/\tilde H)^{\bar \alpha} \cosh^2\theta_5 - (\tilde H/H)^{\bar
  \beta} \sinh^2\theta_5$.

Now in order to obtain the non-susy F/NS5 solution we will apply
the S-duality transformation to \eqn{D1D5D0}.
S-duality will not change the
Einstein frame metric, but will change the dilaton to its inverse. The
RR gauge field $A_{[2]}$ will change to NSNS gauge field and since the S-dual
of D5-brane is NS5-brane which is magnetic we have to take the Hodge dual of
the field-strength of $A_{[6]}$ and that will be an NSNS 3-form. So,
the solution will be given as,
\bea\label{twochargeF}
ds^2 &=& {\hat F}_5^{-\frac{1}{4}} {\hat F}_1^{-\frac{3}{4}}
\left(\frac{H}{\tilde H}\right)^{\frac{\delta_1}{4} + \frac{\delta_2}{2}}
(-dt^2) +
  {\hat F}_5^{-\frac{1}{4}} {\hat F}_1^{\frac{1}{4}}
\left(\frac{H}{\tilde H}\right)^{\frac{\delta_1}{4} + \frac{\delta_2}{2}}
\sum_{i=1}^4 (dx^i)^2 \nn
& & +
  {\hat F}_5^{-\frac{1}{4}} {\hat F}_1^{-\frac{3}{4}}
\left(\frac{H}{\tilde H}\right)^{-\frac{3\delta_1}{4} -
\frac{3\delta_2}{2}} (dx^5)^2 + \left(H\tilde{H}\right)
\left(\frac{H}{\tilde H}\right)^{-\frac{\delta_1}{4} -
\frac{\delta_2}{2}} {\hat F}_5^{\frac{3}{4}} {\hat
F}_1^{\frac{1}{4}}\left(dr^2 + r^2 d\Omega_3^2\right),\nn e^{2\tilde
\phi} &=& {\hat F}_5 {\hat F}_1^{-1}
 \left(\frac{H}{\tilde H}\right)^{-2\delta_1 + 2\delta_2},\nn
B_{[2]} &=& \frac{1}{2}\sinh2\theta_1 \left(\frac{{\hat C}_1} {{\hat
F}_1}\right) dt \wedge dx^5,\nn H_{[3]} &=& b {\rm Vol}(\Omega_3).
\eea Let us now make a coordinate transformation from the radial
coordinate $r$ to $\rho$ as \be\label{rtorho} r = \rho
\left(\frac{1+\sqrt{f}}{2}\right), \qquad {\rm with,} \quad f = 1 -
\frac{4\omega^2}{\rho^2} \equiv 1 - \frac{\rho_0^2}{\rho^2}. \ee
Using \eqn{rtorho} we find $H/\tilde H = f^{-1/2}$. Then in terms of
this new Schwarzschild-like coordinate we can rewrite the solution
\eqn{twochargeF} as follows, \bea\label{twochargeFSch} ds_{\rm
str}^2 &=& e^{\tilde{\phi}/2} ds^2 = G_1^{-1} f^{\frac{{\hat
\alpha}_1}{2}+\frac{\delta_1}{8} - \frac{\delta_2}{2}} (-dt^2) +
f^{\frac{\delta_1}{8} - \frac{\delta_2}{2}} \sum_{i=1}^4 (dx^i)^2
\nn & &\qquad\qquad\quad + G_1^{-1} f^{\frac{{\hat \alpha}_1}{2} +
\frac{5\delta_1}{8} + \frac{\delta_2}{2}} (dx^5)^2 + G_5
f^{-\frac{{\hat \alpha}_5}{2} + \frac{3\delta_1}{8} + \frac{1}{2}}
\left(\frac{d\rho^2}{f} + \rho^2 d\Omega_3^2\right),\nn e^{2\tilde
\phi} &=& G_5 G_1^{-1}
 f^{-\frac{{\hat \alpha}_5}{2} + \frac{{\hat \alpha}_1}{2} +\delta_1
- \delta_2},\nn B_{[2]} &=& \frac{1}{2}\sinh2\theta_1 \left(\frac{1
- f^{\frac {{\hat \alpha}_{1} + {\hat \beta}_{1}}{2}}} {G_1}\right)
dt \wedge dx^5, \qquad H_{[3]} = b {\rm Vol}(\Omega_3). \eea Note
here that we have written the metric in the string frame. The
functions $G_{1,5}$ are defined as, \be\label{gfunctions} G_{1,5} =
{\hat F}_{1,5} f^{\frac{{\hat \alpha}_{1,5}}{2}} =
\cosh^2\theta_{1,5} - f^{\frac{{\hat \alpha}_{1,5} + {\hat
\beta}_{1,5}}{2}} \sinh^2\theta_{1,5} \ee where ${\hat F}_{1,5}$ are
as defined in \eqn{newfunctions}, with $H/\tilde H = f^{-1/2}$. The
parameter relations remain the same as given in \eqn{parametersnew}.
The charge parameter is given as $b = (1/2) ({\hat \alpha}_5 + {\hat
\beta}_5) \rho_0^2 \sinh2\theta_5$. This solution
\eqn{twochargeFSch} represents the non-susy F/NS5 solution and is
characterized by six independent parameters, namely, $({\hat
\alpha}_1 + {\hat \beta}_1)$, $({\hat \alpha}_5 + {\hat \beta}_5)$,
$\rho_0$, $\theta_1$, $\theta_5$ and $\delta_2$. We will get the
two-charge non-susy F-string solution as a special case of this
solution when compactified on T$^4$ and will be discussed later.

From the parameter relations \eqn{parametersnew} it is clear that if we
put
\be\label{blackFNS5parameter1}
\hat{\alpha}_1 + \hat{\beta}_1 = 2,\qquad \hat{\alpha}_5 + \hat{\beta}_5 = 2
\ee
such that the functions
\be\label{newG}
G_{1,5} \rightarrow \G_{1,5} = 1 + \rho_0^2 \sinh^2\theta_{1,5}/\rho^2
\ee
 take the form of the usual harmonic functions and also put
\be\label{blackFNS5parameter2}
\delta_2 = -1/3
\ee
 (which implies $\delta_1=-4/3$, $\hat{\alpha}_1=2$, $\hat{\beta}_1=0$,
 $\hat{\alpha}_5=0$ and $\hat{\beta}_5=2$), then the solution
 \eqn{twochargeFSch} reduces to
 \bea\label{blackFNS5}
ds_{\rm str}^2 &=& \G_1^{-1}\left(-f dt^2 + (dx^5)^2\right) +
\sum_{i=1}^4 (dx^i)^2 + \G_5 \left(\frac{d\rho^2}{f} + \rho^2
d\Omega_3^2\right),\nn e^{2\tilde \phi} &=& \G_5 \G_1^{-1},\nn
B_{[2]} &=& \left(1-\G_1^{-1}\right) \coth\theta_1 dt \wedge dx^5,
\qquad H_{[3]} = b {\rm Vol}(\Omega_3). \eea This is precisely the
black F/NS5 solution. On the other hand, if we put
\be\label{FNS5bubbleparameter} \hat{\alpha}_1 + \hat{\beta}_1 = 2,
\qquad \hat{\alpha}_5 + \hat{\beta}_5 = 2, \qquad {\rm and} \qquad
\delta_2 = 1/3 \ee (which implies $\delta_1=4/3$,
$\hat{\alpha}_1=0$, $\hat{\beta}_1=2$, $\hat{\alpha}_5=2$ and
$\hat{\beta}_5=0$), then the solution \eqn{twochargeFSch} reduces to
\bea\label{FNS5BON} ds_{\rm str}^2 &=& \G_1^{-1}\left(- dt^2 + f
(dx^5)^2\right) + \sum_{i=1}^4 (dx^i)^2 + \G_5
\left(\frac{d\rho^2}{f} + \rho^2 d\Omega_3^2\right),\nn e^{2\tilde
\phi} &=& \G_5 \G_1^{-1},\nn B_{[2]} &=& \left(1-\G_1^{-1}\right)
\coth\theta_1 dt \wedge dx^5, \qquad H_{[3]} = b {\rm
Vol}(\Omega_3). \eea This is the F/NS5 KK BON solution. Here in
order to avoid conical singularity at $\rho=\rho_0$, the coordinate
$x^5$ must be periodic with period \be\label{period} L = 2\pi\rho_0
\cosh\theta_1 \cosh\theta_5. \ee It is therefore clear that
\eqn{twochargeFSch} is the solution which interpolates between the
black or non-extremal F/NS5 solution \eqn{blackFNS5} and the KK BON
solution \eqn{FNS5BON}, by continuously varying the parameters
$\hat{\alpha}_{1,5}$, $\hat{\beta}_{1,5})$ and $\delta_2$
characterizing the solution and there is no need to take the double
Wick rotation.

Similar interpolation from the black D1/D5 configuration to KK BON
can be shown from the general non-susy intersecting D1/D5 system
with chargeless D0 brane solution given in \eqn{D1D5D0}. In order to
show this we will first go to the Schrodinger-like coordinate using
\eqn{rtorho}. In this new coordinate the solution \eqn{D1D5D0} in
the string frame takes the form, \bea\label{D1D5D0Sch} ds_{\rm
str}^2 &=& e^{{\phi}/2} ds^2 = G_1^{-\frac{1}{2}} G_5^{-\frac{1}{2}}
f^{\frac{{\hat \alpha}_1}{4} + \frac{{\hat \alpha}_5}{4} -
\frac{3\delta_1}{8}} (-dt^2) + G_1^{\frac{1}{2}} G_5^{-\frac{1}{2}}
f^{\frac{{\hat \alpha}_5}{4} - \frac{{\hat \alpha}_1}{4} -
\frac{3\delta_1}{8}} \sum_{i=1}^4 (dx^i)^2 \nn & & +
G_1^{-\frac{1}{2}} G_5^{-\frac{1}{2}} f^{\frac{{\hat \alpha}_1}{4} +
\frac{{\hat \alpha}_5}{4} + \frac{\delta_1}{8} + \delta_2} (dx^5)^2
+ G_5^{\frac{1}{2}} G_1^{\frac{1}{2}} f^{-\frac{{\hat \alpha}_1}{4}
-\frac{{\hat \alpha}_5}{4} - \frac{\delta_1}{8} + \frac{\delta_2}{2}
+ \frac{1}{2}} \left(\frac{d\rho^2}{f} + \rho^2
d\Omega_3^2\right),\nn e^{2\phi} &=& G_5^{-1} G_1
 f^{\frac{{\hat \alpha}_5}{2} - \frac{{\hat \alpha}_1}{2} -\delta_1
+ \delta_2},\nn A_{[2]} &=& \frac{1}{2}\sinh2\theta_1 \left(\frac{1
- f^{\frac {{\hat \alpha}_{1} + {\hat \beta}_{1}}{2}}} {G_1}\right)
dt \wedge dx^5, \nn A_{[6]} &=& \frac{1}{2} \sinh2\theta_1
\left(\frac{1 - f^{\frac {{\hat \alpha}_{5} + {\hat \beta}_{5}}{2}}}
{G_5}\right)  dt \wedge dx^1 \wedge dx^2 \wedge dx^3 \wedge dx^4
\wedge dx^5. \eea The parameter relations are as given before in
\eqn{parametersnew}. The various functions appeared in the solution
are as defined earlier. Now again we find that if we choose
\be\label{blackD1D5parameter} {\hat \alpha}_1 + {\hat \beta}_1 = 2,
\quad {\hat \alpha}_5 + {\hat \beta}_5 = 2, \quad {\rm and} \quad
\delta_2 = - \frac{1}{3} \ee implying (from \eqn{parametersnew})
$\delta_1 = -4/3$, ${\hat
  \alpha}_1 = 2$, ${\hat \beta}_1 = 0$ and ${\hat \alpha}_5 = 0$,
${\hat \beta}_5 = 2$, then the solution \eqn{D1D5D0} reduces to,
\bea\label{blackD1D5} ds_{\rm str}^2 &=& \G_1^{-\frac{1}{2}}
\G_5^{-\frac{1}{2}} \left(- f dt^2 + (dx^5)^2\right) +
\G_1^{\frac{1}{2}} \G_5^{-\frac{1}{2}}\sum_{i=1}^4 (dx^i)^2 +
\G_1^{\frac{1}{2}} \G_5^{\frac{1}{2}} \left(\frac{d\rho^2}{f} +
\rho^2 d\Omega_3^2\right),\nn e^{2\phi} &=& \G_5^{-1} \G_1,\nn
A_{[2]} &=& \left(1- \G_1^{-1}\right) \coth\theta_1 dt \wedge
dx^5,\nn A_{[6]} &=&  \left(1-\G_5^{-1}\right) \coth\theta_5 dt
\wedge dx^1 \wedge dx^2 \wedge dx^3 \wedge dx^4 \wedge dx^5. \eea
Note here that for the above choice of parameters
\eqn{blackD1D5parameter}, $G_{1,5} \to
\G_{1,5}=1+\rho_0^2\sinh^2\theta_{1,5}/\rho^2$. This is precisely
the black D1/D5 solution. On the other hand if we choose
\be\label{bubbleparameter} {\hat \alpha}_1 + {\hat \beta}_1 = 2,
\quad {\hat \alpha}_5 + {\hat \beta}_5 = 2, \quad {\rm and} \quad
\delta_2 = \frac{1}{3} \ee implying (from \eqn{parametersnew})
$\delta_1 = 4/3$, ${\hat
  \alpha}_1 = 0$, ${\hat \beta}_1 = 2$ and ${\hat \alpha}_5 = 2$,
${\hat \beta}_5 = 0$, then the metric in \eqn{D1D5D0} reduces to,
\be\label{bubble} ds_{\rm str}^2 = \G_1^{-\frac{1}{2}}
\G_5^{-\frac{1}{2}} \left(- dt^2 + f (dx^5)^2\right) +
\G_1^{\frac{1}{2}} \G_5^{-\frac{1}{2}}\sum_{i=1}^4 (dx^i)^2 +
\G_1^{\frac{1}{2}} \G_5^{\frac{1}{2}} \left(\frac{d\rho^2}{f} +
\rho^2 d\Omega_3^2\right). \ee The other fields remain the same as
in \eqn{blackD1D5}. To avoid the conical singularity at
$\rho=\rho_0$ the periodicity of $x^5$ coordinate remains the same
as in \eqn{period}. This is the corresponding KK BON. Again we see
that by continuously varying the parameters ${\hat \alpha}_{1,5}$,
${\hat \beta}_{1,5}$, and $\delta_2$ the solution \eqn{D1D5D0}
smoothly changes from the black D1/D5 solution to the KK BON.

\section{Initial data analysis}

In the previous section we obtained both non-susy F/NS5
(eq.\eqn{twochargeFSch}) and D1/D5 (eq.\eqn{D1D5D0Sch}) solutions
characterized by six independent parameters. Further we have seen
that when three of the parameters are varied continuously the
solutions nicely interpolate between the corresponding non-extremal
or black solutions and KK BON. Now in order to interpret this
interpolation as a physical transition from the black solution to KK
BON, we must ensure that the final bubble is perturbatively stable
and static such that it does not evolve further. For this purpose we
will perform a time symmetric general bubble initial data analysis.
The time symmetric F/NS5 bubble metric (in Einstein frame) and the
non-trivial dilaton have the forms, \bea\label{initialdata} ds^2 &=&
\G_1^{-\frac{3}{4}} \G_5^{-\frac{1}{4}} f(\rho) (dx^5)^2 +
\G_1^{\frac{1}{4}} \G_5^{-\frac{1}{4}}\sum_{i=1}^4 (dx^i)^2 +
\G_1^{\frac{1}{4}} \G_5^{\frac{3}{4}}\left(\frac{d\rho^2}{f(\rho)
h(\rho)} + \rho^2
  d\Omega_3^2\right),\nn
e^{2\tilde\phi} &=& \G_5 \G_1^{-1} \eea where $\G_{1,5}$ and
$f(\rho)$ were defined earlier and $h(\rho)$ is an unknown function
to be determined from the constraint equation of the time symmetric
initial data. The constraint obtained from the Einstein equation
gives the solution of $h(\rho)$ in the form, \bea && h(\rho)-1\nn &&
= \frac{\lambda\left(\rho^2 + \rho_0^2 \sinh^2\theta_1\right)
              \left(\rho^2 + \rho_0^2 \sinh^2\theta_5\right)}
{\rho^2 \left[3\rho^4 + 2\rho_0^2\rho^2\left(\sinh^2\theta_1 +
\sinh^2\theta_5 -1\right) +
\rho_0^4\left(\sinh^2\theta_1\sinh^2\theta_5 - \sinh^2\theta_1 -
\sinh^2\theta_5\right)\right]}\nn \eea where $\lambda$ is an
integration constant. This therefore gives a four parameter
($\lambda$, $\rho_0$, $\theta_1$ and $\theta_5$) family of time
symmetric, asymptotically flat initial data. Note that the angles
$\theta_1$ and $\theta_5$ are related to the charges associated with
the F-strings and the NS5-branes as follows, \be\label{charges}
Q_{1,5} = \rho_0^2 \sinh2\theta_{1,5}. \ee To avoid the conical
singularity at $\rho=\rho_0$, the length of the $x^5$-circle at
infinity must be \be\label{periodicity} L = 2\pi \rho_0
\cosh\theta_1 \cosh\theta_5 \left(1 +
  \frac{\lambda}{\rho_0^2}\right)^{-\frac{1}{2}}.
\ee
The ADM mass of the bubble can be obtained from the metric in
\eqn{initialdata} as,
\bea\label{ADMmass}
M &=& \frac{\Omega_3 \rho_0^2}{2\kappa^2}\left[\sqrt{1 +
    \left(\frac{Q_1}{\rho_0^2}\right)^2} + \sqrt{1 +
    \left(\frac{Q_5}{\rho_0^2}\right)^2}\right.\nn
& & \qquad\qquad \left. - \frac{1}{4}
\left(\frac{2\pi\rho_0}{L}\right)^2 \left(1 + \sqrt{1 +
\left(\frac{Q_1}{\rho_0^2}\right)^2}\right) \left(1 + \sqrt{1 +
\left(\frac{Q_5}{\rho_0^2}\right)^2}\right)\right] \eea where
$\Omega_3 = 2\pi^2$ is the volume of a unit 3-sphere and $2\kappa^2
= 16\pi G$, with $G$, the Newton's constant. Here
(\ref{periodicity}) has also been used in eliminating the parameter
$\lambda$. Note that for given $Q_{1,5}$ and $L$, the mass
\eqn{ADMmass} takes the value $(\pi^2/\kappa^2)(Q_1 + Q_5 - \pi^2
Q_1 Q_5/L^2)$ for $\rho_0 \to 0$ and for $\rho_0 \to \infty$, $M \to
-\infty$. So, there is no lower bound on mass and the positive
energy theorem fails as was also noticed in \cite{Horowitz:2005vp}.
The mass \eqn{ADMmass} has extremum when \be\label{firstderivative}
\frac{dM}{d\rho_0} = 0 = \frac{\Omega_3 \rho_0}{\kappa^2}
\left(\frac{1} {\cosh2\theta_1} + \frac{1}{\cosh2\theta_5}\right)
\left(1 - \frac{4\pi^2\rho_0^2}{L^2}
\cosh^2\theta_1\cosh^2\theta_5\right) \ee and this gives,
\be\label{extrememass} L = 2\pi\rho_0\cosh\theta_1 \cosh\theta_5.
\ee Comparing this with \eqn{periodicity}, we find that the extremum
occurs at $\lambda = 0$ or $h(\rho) = 1$. The resulting metric in
\eqn{initialdata} is now the spatial part of the static bubble one
obtains from the double Wick rotation of the black F/NS5 solution
\eqn{blackFNS5} in Einstein frame.

Now we will try to see whether the extremum is a local maximum or a
local minimum by evaluating the double derivative $d^2 M/d \rho_0^2$
and determining its sign. We find, \bea\label{secondderivative} &&
\left. \frac{2\kappa^2}{\Omega_3} \frac{d^2 M}
{d\rho_0^2}\right|_{\lambda = 0}\nn && = \frac{16\pi^2
\rho_0^2}{L^2} \cosh2\theta_1 \cosh2\theta_5\left(\frac{1}
{\cosh2\theta_1} + \frac{1}{\cosh2\theta_5}\right)\left(1-\frac{1}
{\cosh2\theta_1} - \frac{1}{\cosh2\theta_5}\right). \eea However, it
is clear that it is not easy to determine the position and the
nature of the extremum from \eqn{extrememass} and
\eqn{secondderivative} along with (\ref{charges}) for given $Q_1,
Q_5$ and $L$. So, we will look at the $M$-$\rho_0$ relation
\eqn{ADMmass} more closely and try to find them in an indirect way.

For this purpose let us define two functions, based on
(\ref{charges}) and (\ref{extrememass}), as follows,
\bea\label{C1C5}  \frac{Q_5}{L^2} &=& \frac{\sinh\theta_5}{2\pi^2
  \cosh\theta_5\cosh^2\theta_1} = \frac{\sinh\theta_5}{\pi^2 (1+\sqrt{1 + k^2
\sinh^2 2\theta_5})\cosh\theta_5} \equiv C_5(\theta_5), \nn
\frac{Q_1}{L^2} &=& \frac{\sinh\theta_1}{2\pi^2
  \cosh\theta_1\cosh^2\theta_5} = \frac{\sinh\theta_1}{\pi^2 (1+\sqrt{1 +
k^{-2} \sinh^2 2\theta_1})\cosh\theta_1} \equiv C_1(\theta_1)  \eea
where $k=Q_1/Q_5 = \sinh2\theta_1/\sinh2\theta_5 \neq 0$. Note that
for given $Q_5$, $Q_1$ and $L$ (therefore given $k$), the solution of
either $C_5(\theta_5)$ or $C_1(\theta_1)$ equation (they are
correlated by $k$) above will give us the position of $\rho_0$ where
the extremum occurs. This is because once we obtain $\theta_5$ and
$\theta_1$, $\rho_0$ where extremum occurs can be determined from
\eqn{charges} or \eqn{extrememass}. Now to solve \eqn{C1C5} we first
note that $C_5(\theta_5)$ approaches zero for both $\theta_5 \to 0$
and $\theta_5 \to \infty$, while in between it is non-zero and
positive. So, we expect at least one maximum for $C_5(\theta_5)$ for
some $0<\theta_5<\infty$. The same is true for $C_1(\theta_1)$. In
the following we will argue that there exists only one such maximum
for either $C_5(\theta_5)$ or $C_1(\theta_1)$ (the two are
correlated by $C_1 (\theta_1) = k C_5 (\theta_5)$). For concreteness
we will focus on $C_5(\theta_5)$ only and give results for
$C_1(\theta_1)$. Differentiating $C_5(\theta_5)$ in \eqn{C1C5} with
respect to $\theta_5$ and putting it to zero we get, \be\label{dC5}
\frac{dC_5(\theta_5)}{d\theta_5} = 0 \Rightarrow 16k^2 x^4 + 16 k^2
x^3 - 4x -1 = 0\ee where $x=\sinh^2\theta_5 \geq 0$. Let us define
$G(x) = 16k^2 x^4 + 16 k^2 x^3 - 4 x -1$. Then by looking at its
behavior we can infer that there is a unique solution of the
equation $G(x) = 0$. First note that $G(x=0)=-1$ and $G(x\to
\infty)\to \infty$. Further, $d^2 G/dx^2 = 96k^2x(2x+1)>0$ for all
$x>0$ which means that $G(x)$ has a unique minimum in
$0 < x < \infty$. Now since
$G(x=0) = -1 <0$, $G(x)$ will cut the $x$-axis only once.
So, it is clear that there exists only one $\theta_5$
(let us call this $\theta_{\rm 5 max}$) for which $G(x) = 0$. This
$\theta_{\rm 5 max}$ must give the maximum of $C_5(\theta_5)$,
denoted as $C_{\rm 5max}$. So, given $Q_5$ and $L$ if
$Q_5/L^2>C_{\rm 5 max}$, then there exists no static bubble, if
$Q_5/L^2 = C_{\rm 5 max}$, the ADM mass has a turning point where
the first and the second derivatives vanish\footnote{One can check
that $G (x) = 0$ corresponds precisely to the vanishing second
derivative.} , while if $Q_5/L^2 < C_{\rm 5 max}$, there exist two
static bubbles. In the last case, the solution with large
$\theta_5$, denoted as $\theta_{5\ell}$, will give small $\rho_0$,
while the solution with small $\theta_5$, denoted as $\theta_{5s}$,
will give large $\rho_0$ (see eq.\eqn{charges}). So, we have
$\theta_{5s}<\theta_{\rm 5 max}<\theta_{5\ell}$. Note that since ADM
mass (eq.\eqn{ADMmass}) $M \to -\infty$ as $\rho \to \infty$, we
expect large $\rho_0$ ($\theta_{5s}$) corresponds to the maximum of
mass (or unstable static bubble) while small $\rho_0$
($\theta_{5\ell}$) corresponds to the minimum of mass (or stable
static bubble).

\vspace{.5cm}

\noindent{\it Values of $C_{\rm max}$ and
$\theta_{\rm max}$}

\vspace{.2cm}

Now let us make some estimate of the values of $C_{\rm 5,1\,max}$ and
$\theta_{\rm 5,1\,max}$ discussed above. This will certainly depend on the
value of the parameter $k=Q_1/Q_5=\sinh2\theta_1/\sinh2\theta_5$. So, we
will consider three different cases:
$ (i)\,\, k \sim 1,\,\, (ii)\,\, k \ll 1$ and $ (iii)\,\, k \gg 1$.

\vspace{.2cm}

\noindent{\it Case (i):} When $k \sim 1$, we can make an order of
magnitude estimate of $C_{\rm max}$ and $\theta_{\rm max}$ by taking
$k=1$ for simplicity. In this case eq.\eqn{dC5} can be factorized
as, $(2x-1)(2x+1)^3 =0$, which can be solved to give $x=1/2$ (note
that $x\geq 0$). This in turn gives, $\sinh\theta_{\rm 5 max} =
1/\sqrt 2$ and so, we get, \bea\label{thetaC1} e^{\theta_{\rm 5max}}
&=& \frac{1+\sqrt{3}}{\sqrt{2}} \approx 1.94,\nn C_{\rm 5max} &=&
\frac{1}{3\sqrt{3} \pi^2} \approx 0.02. \eea As we have argued, in
order to have a stable static bubble we need to have $Q_5/L^2 <
C_{\rm 5max} \sim 0.02$ which is a small but fixed number. Also as
$k\sim 1$, so, both $\theta_{\rm 1max}$ and $C_{\rm 1max}$  of the
strings should be of the same order of magnitude as the
corresponding quantities of NS5-branes given in \eqn{thetaC1}.

\vspace{.2cm}

\noindent{\it Case (ii):} When $k \ll 1$, eq.\eqn{dC5} simplifies to
$16k^2(x^4+x^3) = 1 + 4x$. It can be solved next to leading order as
$x = (2k)^{-2/3}(1-(2k)^{2/3}/4)$, which in turn gives, $\sinh\theta_{\rm
  5max} = (2k)^{-1/3} (1 - (2k)^{2/3}/8)$. We, therefore, have
\bea\label{thetaC2} e^{\theta_{\rm 5max}} &=&
\frac{2}{(2k)^{\frac{1}{3}}}\left(1 +
\frac{(2k)^{\frac{2}{3}}}{8}\right),\nn C_{\rm 5max} &=&
\frac{1}{2\pi^2} \left(1 - \frac{3}{4}(2k)^{\frac{2}{3}} \right)
\approx 0.05. \eea So, we now have a larger $C_{\rm 5max}$ and a
very large $\theta_{\rm
  5max}$. Therefore, in this case the condition $Q_5/L^2\ll C_{\rm 5max}$
(which is necessary for the existence of stable static bubble as well as
the occurrence of closed string tachyon condensation as we will see) can
be easily satisfied. From \eqn{C1C5} we find that $\sinh2\theta_{\rm 1max}
= k \sinh2\theta_{\rm 5max} = (2k)^{1/3}$, which is vanishingly small for
$k \ll 1$. This
therefore tells us that even though black NS5-branes do not make transition
to the stable static bubble via closed string tachyon condensation, addition
of a few strings makes this transition possible.

\vspace{.2cm}

\noindent{\it Case (iii):} In this case eq.\eqn{dC5} gives us the solution for
$x$ to the leading order as $x = 1/(2k)^{2/3}$. This implies $\sinh\theta_{\rm
  5max} = 1/(4k)^{1/3}$. Therefore, we have,
\bea\label{thetaC3} e^{\theta_{\rm 5max}} &=& 1 +
\frac{1}{(4k)^{\frac{1}{3}}},\nn C_{\rm 5max} &=& \frac{1}{2\pi^2
k}. \eea So, $C_{\rm 5max}$ is vanishingly small for $k \gg 1$ and
as a result, it might seem that there is no possibility of a
transition from black brane to stable static bubble (also the
occurrence of closed string tachyon condensation) in this case.
However, looking at eq.\eqn{C1C5} we find that $\sinh2\theta_{\rm
1max} = k \sinh2\theta_{\rm 5max} = (4k)^{2/3}/2 \gg 1$. So, we have
very large $\theta_{\rm 1max}$ and also to leading order $C_{\rm
1max} = 1/(2\pi^2) \approx 0.05$. This is good for the transition to
the stable static bubble and the occurrence of closed string tachyon
condensation. This case is just opposite to the previous case in the
sense that strings in the previous case plays the role of NS5-branes
and the NS5-branes in the previous case plays the roles of strings
in this case. So, here, even though the black strings do not make a
transition to stable static bubble via closed string tachyon
condensation, addition of a few NS5-branes makes the transition
possible.

\vspace{.5cm}

\noindent{\it Transition to stable static bubble}

\vspace{.2cm}

After having some estimate on the parameters $C_{\rm 5,1max}$ and
$\theta_{\rm 5,1max}$, we will try to see under what condition the black
configuration will make a transition to stable static bubble (when the
parameter $\theta_{5,1}$ takes values $\theta_{5,1\ell}$ corresponding to
small bubble) and what is the estimate for the values of $\theta_{5,1\ell}$.
This will be necessary for the interpretation of the interpolating solution
as the process of closed string tachyon condensation and will be discussed in
the next section.

We have seen that for $k\sim 1$, when $Q_5/L^2 \sim Q_1/L^2 < C_{\rm
5max} \sim C_{\rm 1max} \approx 0.02$, there exist two static
bubbles, of which the smaller one (smaller $\rho_0$ which
corresponds to bigger $\theta$ i.e. $\theta_{5,1\ell}$) corresponds
to the stable static bubble. Further, the closed string tachyon
condensation requires (as we will discuss later) $Q_1/L^2 \sim
Q_5/L^2 \ll 0.02$ and so, we get from \eqn{C1C5},
\be\label{thetaell1} e^{\theta_{1\ell}} \sim
\sqrt{\frac{2}{Q_5}}\frac{L}{\pi}, \qquad e^{\theta_{5\ell}} \sim
\sqrt{\frac{2}{Q_1}}\frac{L}{\pi}. \ee

On the other hand, for $k \ll 1$, when $Q_5/L^2 < C_{\rm 5max}
\approx 0.05$, again there exist two static bubbles of which
$\theta_{5\ell}$ gives the stable one. In this case since
$\theta_{\rm 5max}$ is large, it will be easier to realize the
closed string tachyon condensation without much requirement on
$Q_5/L^2$ unlike in the previous case. Note that once the condition
$Q_5/L^2 < C_{\rm 5max}$ is satisfied, the corresponding condition
for the strings should also be satisfied automatically. However, now
$\theta_{\rm 1max}$ is vanishingly small, so, let us discuss the
values of the string parameters in a slightly more detail. We know
that \be\label{thetaell2} e^{\theta_{5\ell}} > e^{\theta_{\rm 5max}}
\approx 2(2k)^{-\frac{1}{3}} \ee is very large while $\theta_{1\ell}
> \theta_{\rm 1max} \approx (2k)^{1/3}/2 \ll 1$ can be still very
small or of order one or large depending on how large
$\theta_{5\ell}$ is. One thing is clear that in order to satisfy
eq.\eqn{thetaell2} $k^2\sinh^2 2\theta_{5\ell} >
e^{-2\theta_{5\ell}}$. So, let us consider three different
situations depending on the value of $k^2\sinh^2 2\theta_{5\ell}$
for which $\theta_{1\ell}$ could be either very small or of order
one or very large, namely, (a) $e^{-2\theta_{5\ell}} < k^2 \sinh^2
2\theta_{5\ell} \ll 1$, (b) $k^2\sinh^2 2\theta_{5\ell}=A\sim 1$,
(c) $k^2\sinh^2 2\theta_{5\ell} \gg 1$.

Let us consider (a) and set \be\label{thetaell3} k^2\sinh^2
2\theta_{5\ell} \approx \frac{k^2}{4} e^{4\theta_{5\ell}} = B
e^{-2\theta_{5\ell}}, \ee where $B$ is a parameter whose value will
be determined from the condition \eqn{thetaell2}. From
\eqn{thetaell3} we obtain \be\label{thetaell4} e^{\theta_{5\ell}} =
2^{2/3} B^{1/6} (2k)^{-1/3} \gg 1. \ee Now it can be checked that
for \eqn{thetaell2} to be satisfied the parameter $B$ must be
restricted as, \be\label{Bparameter}
 B > 4.
\ee
 It is not difficult to see that the condition \eqn{Bparameter} on the
$B$-parameter is just another representation of the condition $Q_5/L^2<C_{\rm
   5max}$ for the existence of static bubble. In fact, if we use the latter
condition we can recover the restriction on the $B$-parameter as follows.
 Let us first rewrite $Q_5/L^2$ given in \eqn{C1C5} using \eqn{thetaell3}
and then use \eqn{thetaell4} to get,
\be\label{Q5byL2}
\frac{Q_5}{L^2} = \frac{1}{2\pi^2}\left[1 -
  \frac{1+\frac{B}{8}}{(2B)^{\frac{1}{3}}} (2k)^{\frac{2}{3}}\right] < C_{\rm
  5max} \approx
  \frac{1}{2\pi^2}\left[1-\frac{3}{4}(2k)^{\frac{2}{3}}\right],
\ee where in the last step we have used \eqn{thetaC2}. We thus get
from \eqn{Q5byL2}, $(4+B/2)/3 > (2B)^{1/3}$. This is consistent only
if $B=4(1+\epsilon)$ with $\epsilon>0$. Now $\theta_{1\ell}$ can be
determined from the following, \be\label{theta1ell1}
\sinh2\theta_{1\ell} = k \sinh2\theta_{5\ell} \approx \frac{k}{2}
e^{2\theta_{5\ell}} = \frac{1}{2} (4Bk)^{\frac{1}{3}} \ll 1, \ee
where in the last inequality we have used $Bk \ll 2$ which follows
from $B e^{- 2 \theta_{5l}} \ll 1$, \eqn{thetaell2} and $B > 4$. So,
\eqn{theta1ell1} implies small $\theta_{1\ell}$ which has the form,
\be\label{theta1ell2} \theta_{1\ell} =
\frac{1}{4}(4Bk)^{\frac{1}{3}} \ll 1. \ee

We now come to situation (b), where $k^2\sinh^2 2\theta_{5\ell} = A
\sim 1$. In this case we can solve \eqn{C1C5} to obtain
$A=[L^4/(\pi^4 Q_5^2)] (1-2\pi^2 Q_5/L^2)$. This can now be used to
obtain \be\label{theta1ell5} e^{\theta_{5\ell}} = \frac{2L}{\pi
\sqrt{Q_5}} \left(1-\frac{2\pi^2 Q_5}{L^2}\right)^{\frac{1}{4}}
(2k)^{-\frac{1}{2}} \gg 1. \ee We therefore have $\theta_{1\ell}$
as,
 \be\label{theta1ell3}
 \sinh2\theta_{1\ell} = k \sinh2\theta_{5\ell} = \frac{k}{2}
 e^{2\theta_{5\ell}} = \frac{L^2}{\pi^2 Q_5}\left(1-\frac{2\pi^2
    Q_5}{L^2}\right)^{\frac{1}{2}} = {\rm finite}
\ee
from which we can now solve to get
\be\label{theta1ell4}
e^{\theta_{1\ell}} = \frac{L}{\pi \sqrt{Q_5}} \left(1- \frac{2\pi^2 Q_5}{L^2}
\right)^{\frac{1}{4}} \left[1 + \sqrt{1+\frac{\pi^4
      Q_5^2}{L^4}\left(1-\frac{2\pi^2
        Q_5}{L^2}\right)^{-1}}\right]^{\frac{1}{2}}.
\ee

Next we consider situation (c) for which $k^2 \sinh^2
2\theta_{5\ell} \gg 1$. In this case we get from \eqn{C1C5},
\be\label{largetheta1} \frac{Q_5}{L^2} \approx \frac{1}{\pi^2 k
\sinh2\theta_{5\ell}} \approx \frac{2}{\pi^2 k
  e^{2\theta_{5\ell}}} \quad \Rightarrow \quad e^{\theta_{5\ell}} = \frac{2L}
{\pi \sqrt{Q_5}} (2k)^{-\frac{1}{2}} \gg 1. \ee Therefore from
$\sinh2\theta_{1\ell} = k \sinh 2\theta_{5\ell} = L^2/(\pi^2 Q_5)
\gg 1$, we have \bea\label{largetheta2} e^{\theta_{1\ell}} &=&
\frac{L}{\pi \sqrt{Q_5}}\left[1+\sqrt{1+\frac{\pi^4
Q_5^2}{L^4}}\right]^{\frac{1}{2}}\nn &\approx& \sqrt{\frac{2}{Q_5}}
\frac{L}{\pi}\left(1+\frac{\pi^4
    Q_5^2}{8L^4}\right) \gg 1
\eea
We thus conclude that for $k\ll 1$, stable static bubble can be obtained
but in this case even though
$\theta_{5\ell}$ is always large $\theta_{1\ell}$ could be small, large or of
order 1 depending on how large the value of $\theta_{5\ell}$ is.

The case of $k \gg 1$ can be discussed in a similar fashion if we set
$k'=1/k$. Then here $\theta_{1\ell}$ takes the role of $\theta_{5\ell}$ in
the previous case and vice-versa. So, we will not repeat the discussion here.
Thus we have seen how the initial data analysis helps us to discuss the
stability of the final bubble in various cases.

\section{Physical interpretation of the interpolations}

In the previous section we have seen by initial data analysis that under
certain circumstances, the black F/NS5 or D1/D5 can indeed make a transition
to locally stable static bubbles. In these cases the interpolating solutions
described
in section 2 make sense as the final bubbles are classically stable and do not
evolve further. In this section we will try to give a physical interpretation
to these interpolations as the perturbative stringy process of closed string
tachyon condensation. As we will see if the closed string tachyon condensation
is the possible mechanism for the transition from black solution to KK BON,
several conditions have to be satisfied. We will discuss non-susy F/NS5 case
first and then the case of non-susy D1/D5.

\subsection{Non-susy F/NS5}

The classical supergravity solution interpolating between black
F/NS5 and KK BON is given in \eqn{twochargeFSch}. This interpolation
can be regarded as a transition from black F/NS5 to KK BON under
certain conditions. Since this is a topology changing transition
like what happens for stringy process of closed string tachyon
condensation, the latter can be taken as a possible mechanism for
the transition as we will argue. We will consider the closed string
tachyon condensation to occur on the horizon, where the space-time
curvature must be small compared to the string length otherwise the
supergravity description will break down. The black F/NS5
supergravity configuration is given in \eqn{blackFNS5} where the
coordinate $x^5$ is compact. In order to have a closed string
tachyon condensation the size of the $x^5$ circle must satisfy
\be\label{LblackFNS5} L = l_s \cosh\theta_1 \ee where $l_s$ is the
fundamental string length. The size of the horizon and the two
charges associated with the solution can be written as,
\be\label{ZblackFNS5} Z = \rho_0 \cosh\theta_5, \qquad Q_1 =
\rho_0^2 \sinh2\theta_1, \qquad Q_5 = \rho_0^2 \sinh2\theta_5. \ee
The corresponding KK BON solution is given in \eqn{FNS5BON}. We have
found that to avoid conical singularity at $\rho_0$, $x^5$ must be
periodic with a periodicity given in \eqn{period}. We denote the
various bubble quantities with a subscript `$b$' and also, since the
parameters for the two solutions need not be the same, we denote the
bubble parameters with a `tilde'. So, the periodicity, size of the
bubble, and the fluxes can be written as follows:
\be\label{LZFNS5BON} L_b = 2\pi \rh_0 \cosh \thet_1 \cosh \thet_5,
\quad Z_b = \rh_0 \cosh \thet_5, \quad Q_{1b} = \rh_0^2 \sinh
2\thet_1, \quad Q_{5b} = \rh_0^2 \sinh 2\thet_5. \ee Now if black
F/NS5 solution makes a transition to KK BON, then the quantities
given above for the two solutions must be equated. In fact we must
have $Q_1= Q_{1b}$, $Q_5=Q_{5b}$ as exact relations due to the
charge conservation. Also we must have $L=L_b$ as exact relation as
well since this is the asymptotic radius of $x^5$-circle and tachyon
condensation occurs on the horizon as a local process. However, the
radius of the horizon and the bubble size will have only their
leading order (i.e., the classical part) equal since the tachyon
condensation occurs on the horizon and there may be some quantum
corrections. This is evident that $\theta_5$ and
$\tilde\theta_5$ always remain large for the transition as will be
demonstrated in the following and the ratio of the sub-leading order
over the leading order is of the order of ${\cal O} (e^{- 2
\theta_5} \,{\rm or}\, e^{- 2\tilde\theta_5})$, i.e.,  exponentially
small. For this reason, we can have only $Z \approx Z_b \gg l_s$.

It is clear from \eqn{LblackFNS5}, that $\theta_1$ is very large and
so, we have, \be\label{thetaone} e^{\theta_1} \approx
\frac{2L}{l_s}. \ee Also from the definition of $k$ as well as
\eqn{ZblackFNS5} we have \be\label{thetafiverhozero} \sinh 2\theta_5
\approx \frac{2L^2}{l_s^2 k}, \qquad \rho_0 \approx
\sqrt{\frac{kQ_5}{2}} \frac{l_s}{L}. \ee Now let us first assume $k
\sim 1$, implying that $\theta_5$ is also very large. So, the
$\theta_5$ equation in \eqn{thetafiverhozero} simplifies to
$e^{\theta_5} \approx 2L/(\sqrt {k} l_s)$. Further, from
$Q_5=Q_{5b}$ and $Z \approx Z_b$ we get $\tanh\theta_5 \approx \tanh
\thet_5$, which implies $\thet_5$ is very large and so, $\thet_1$ is
also very large. Moreover, using these as well as $L = L_b \gg
l_s,\, Q_1 = Q_{1b}$ in addition, we  have, \be\label{condition1}
e^{\theta_1} \gg e^{\thet_1}, \quad \rho_0 \ll \rh_0, \quad
e^{\theta_5} \gg e^{\thet_5}, \quad {\rm and} \quad \rho_0
e^{\theta_5} \approx \rh_0 e^{\thet_5} \gg l_s. \ee We already
discussed that for $k\sim 1$, the stable static bubble corresponds
to (see eq.\eqn{thetaell1}) \be\label{bubbletheta} e^{\thet_{1\ell}}
\sim \sqrt{\frac{2}{Q_5}}\frac{L}{\pi}, \qquad e^{\thet_{5\ell}}
\sim \sqrt{\frac{2}{k Q_5}}\frac{L}{\pi} \ee with $Q_5/L^2 \ll
C_{\rm 5max} \approx 0.02$ and from \eqn{LZFNS5BON}, we have
\be\label{bubblerho} \rh_0 = \pi \sqrt{k Q_5}
\sqrt{\frac{Q_5}{L^2}}. \ee Given $\sqrt{Q_5} \gg l_s$, along with
black F/NS5 parameters \eqn{thetaone} and \eqn{thetafiverhozero} and
also the bubble parameters \eqn{bubbletheta}, \eqn{bubblerho}, it is
obvious that all the conditions in \eqn{condition1} can be
satisfied. So, the transition can indeed be caused by the closed
string tachyon condensation.

Next we consider $k \ll 1$ case. Here also both $\theta_1$ and
$\theta_5$ are quite large and so as in the previous case
\eqn{thetaone} and \eqn{thetafiverhozero} with $\sinh2\theta_5$
replaced by $e^{2\theta_5}/2$ hold. As before in this case also
$\thet_{5\ell}$ is always large (which follows from $Q_5 = Q_{b5}$
and $Z \approx Z_b$), but $\thet_{1\ell}$ can be small, of order one
or large, as we discussed earlier, depending on how large the value
of $\thet_{5\ell}$ is. For small $\thet_{1\ell}$, we have from
\eqn{thetaell4} and \eqn{theta1ell2} \be\label{bubbletheta1}
e^{\thet_{5\ell}} = 2^{\frac{2}{3}} B^{\frac{1}{6}}
(2k)^{-\frac{1}{3}}, \qquad \thet_{1\ell} = \frac{1}{4}
(4Bk)^{\frac{1}{3}} \ee and from \eqn{LZFNS5BON} we have,
\be\label{bubblerho1} \rh_0 = (2B)^{-\frac{1}{6}} \sqrt{Q_5}
(2k)^{\frac{1}{3}}. \ee The condition for the closed string tachyon
condensation in this case is slightly different (since
$\thet_{1\ell}$ is small) from the previous case \eqn{condition1}
as, \be\label{condition2} e^{\theta_1} \gg 4\pi \cosh \thet_1,\quad
\rho_0 \ll \rh_0, \quad e^{\theta_5} \gg e^{\thet_5}, \quad {\rm
and}\quad \rho_0 e^{\theta_5} \approx \rh_0 e^{\thet_5} \gg l_s .
\ee It can be verified that all these conditions can be satisfied
using \eqn{thetaone} and \eqn{thetafiverhozero} for black F/NS5 and
\eqn{bubbletheta1} and \eqn{bubblerho1} for the bubble provided \be
\sqrt{Q_5} \gg l_s, \ee which can be satisfied easily.

For the case of finite $\thet_{1\ell}$, as discussed earlier (see
\eqn{theta1ell5} and \eqn{theta1ell4}), we now have
\bea\label{bubbletheta2}
 e^{\thet_{5\ell}} &=&
\frac{2L}{\pi \sqrt{Q_5}} \left(1-\frac{2\pi^2
Q_5}{L^2}\right)^{\frac{1}{4}} (2k)^{-\frac{1}{2}}, \nn
\sinh2\thet_{1\ell} &=& \frac{L^2}{\pi^2 Q_5}\left(1-\frac{2\pi^2
    Q_5}{L^2}\right)^{\frac{1}{2}},
\eea
also from \eqn{LZFNS5BON} we have
\be\label{bubblerho2}
\rh_0 = \frac{\pi}{\sqrt{2}}\frac{Q_5}{L}\left(1 - \frac{2\pi^2
    Q_5}{L^2}\right)^{-\frac{1}{4}} (2k)^{\frac{1}{2}}.
\ee
The conditions for the closed string tachyon condensation in this case remain
the same as in \eqn{condition2}. It can be checked that they can be satisfied
if we use \eqn{bubbletheta2} and \eqn{bubblerho2} provided
$\sqrt{Q_5} \gg l_s$.

For the case of large $\thet_{1\ell}$, we have from
\eqn{largetheta1} and \eqn{largetheta2} \bea e^{\thet_{5\ell}} &=&
\frac{2L} {\pi \sqrt{Q_5}} (2k)^{-\frac{1}{2}}, \nn
e^{\thet_{1\ell}} &=& \sqrt{\frac{2}{Q_5}}
\frac{L}{\pi}\left(1+\frac{\pi^4
    Q_5^2}{8L^4}\right)
\eea with $L^2/(\pi^2 Q_5) \gg 1$. Then from \eqn{LZFNS5BON} we get,
\be \rh_0 = \sqrt{Q_5}\frac{\pi \sqrt{Q_5}}{L} (2k)^{\frac{1}{2}}.
\ee Here the conditions for the closed string tachyon condensation
\eqn{condition1} can be satisfied provided $\sqrt{Q_5} \gg l_s$.

We would like to point out that since the occurrence of closed
string tachyon condensation requires $\rho_0 \ll \rh_0$, for the
supergravity description to remain valid, the string coupling must
be small at $\rho=\rho_0$. Putting in the asymptotic string
coupling, the dilaton in this case has the form $e^{2\tilde\phi} =
g_s^2 \G_5/\G_1$ and so, its value at $\rho=\rho_0$ is $g_s^2
\cosh^2\theta_5/ \cosh^2\theta_1 \approx g_s^2 e^{2\theta_5 -
2\theta_1} \approx g_s^2 Q_5/Q_1$. Note that both $\theta_1$ and
$\theta_5$ are large here and so, the hyperbolic functions can be
approximated by the exponentials. For F/NS5 solution we can write
$Q_1 = (N_1/V_4) \alpha'^3 g_s^2$, and $Q_5 = N_5 \alpha'$, where
$V_4$ is the volume covered by NS5-branes transverse to F-strings
and $\alpha'=l_s^2$. If we introduce a dimensionless volume by $v_4
= V_4/\alpha'^2$, then $e^{2\tilde\phi} = v_4 N_5/N_1$ and so, for
this to remain small, an additional condition has to be satisfied
which is $v_4 N_5 \ll N_1$ and if we have $v_4 N_5 \gg N_1$, we must
go to the S-dual D1/D5 configuration which we will discuss in the
next subsection.

Now we come to the case of $k \gg 1$. In this case the string charge
dominates. The black F/NS5 parameters remain the same as given in
\eqn{thetaone} and \eqn{thetafiverhozero}. It is now more proper to
use string charge $Q_1 = k Q_5$, instead of $Q_5$. Note that both
$Q_1$ and $Q_5$ are supposed to be fixed macroscopic quantities and
so $k \gg 1$ implies that $k$ is also supposed to be large but
fixed. As discussed in the last paragraph, $Q_1 = k Q_5$ gives
$g_s^2 N_1 /v_4 = k N_5$. Since $g_s \ll 1$ and $v_4 \ge 1$, large
$k$ requires very large $N_1 /v_4$ which is surely more difficult to
realize. On the other hand, $k = \sinh 2\theta_1 /\sinh 2\theta_5
\approx e^{2\theta_1}/(2 \sinh 2\theta_5)$ (since $\theta_1$ is
large, being independently determined by \eqn{thetaone}). So the
largeness of $k$ depends on the value of $\theta_5$ which can now be
small, of order unity and large. Therefore a rather large $\theta_5$
can give rise to a fixed large $k$ which should be easier to
realize. While the other two scenarios can also be considered, we
here limit ourselves to the large $\theta_5$ for the reason just
given and for simplicity as well.  For now, the closed string
tachyon condensation condition is again given by \eqn{condition2}.
Note that $\thet_5$ continues to be large (due to $\tanh\theta_5
\approx \tanh\tilde\theta_5$). On the bubble side, we can repeat the
earlier analysis by using string rather than fivebrane, but by
replacing $k \to k'=1/k \ll 1$. Now $\thet_{1\ell}$ is large, but
$\thet_{5\ell}$ can be small, of order one or large. As just
mentioned only large $\thet_{5\ell}$ is relevant here. What we have
done for $k \ll 1$ cases can be borrowed here with the replacements
$1 \leftrightarrow 5$ and $k \to 1/k$.
 With this, we have now on the
bubble side, \bea e^{\thet_{1\ell}} &=& \frac{2L}{\pi
  \sqrt{Q_1}}\left(\frac{k}{2}\right)^{\frac{1}{2}},\nn
e^{\thet_{5\ell}} &=& \sqrt{\frac{2}{Q_1}} \frac{L}{\pi}
\left(1+\frac{\pi^4 Q_1^2}{8L^4}\right) \eea with now $L^2/(\pi^2
Q_1) \gg 1$. Then from \eqn{LZFNS5BON} we get, \be \rh_0 =
\sqrt{Q_1} \frac{\pi \sqrt{Q_1}}{L}
\left(\frac{2}{k}\right)^{\frac{1}{2}}. \ee The condition
\eqn{condition2} can again be satisfied provided $\sqrt{Q_5} \gg
l_s$. This again requires that there be enough number of NS5-branes
present even though $Q_1/Q_5 = k \gg 1$. In other words, so long as
there are enough number of NS5-branes present, this condition can be
satisfied.

In this case since $\theta_1$ is much greater than $\theta_5$, the dilaton
will always remain small here and there is no additional condition to be
satisfied.

In summary, we have seen that with the presence of F-strings in NS5-brane
systems or the presence of NS5-branes in F-string systems makes the transition
from the black solution to KK BON possible via the stringy closed string
tachyon condensation process.

\subsection{Non-susy D1/D5}

The classical supergravity solution interpolating between black
D1/D5 and KK BON is given in \eqn{D1D5D0Sch}. As we have seen in
section 3 by initial data analysis, this interpolation can be
regarded as a transition from black D1/D5 to KK BON under certain
conditions. Note that although we have not explicitly performed the
initial data analysis for non-susy D1/D5 solution, but since this is
S-dual to F/NS5 solution, the analysis remains the same. Again we
will try to give a physical interpretation of this transition as the
closed string tachyon condensation. As before, we will consider the
closed string tachyon condensation to occur on the horizon, where
the space-time curvature must be small compared to the string length
otherwise the supergravity description will break down. The black
D1/D5 supergravity configuration is given in \eqn{blackD1D5} where
the coordinate $x^5$ is compact. In order to have a closed string
tachyon condensation the size of the $x^5$ circle must satisfy
\be\label{LblackD1D5} L = \bar{l}_s \cosh^{\frac{1}{2}}\theta_1
\cosh^{\frac{1}{2}}\theta_5. \ee Here we have put a `bar' on $l_s$
to distinguish that from the previous section and they are related
by S-duality. The size of the horizon and the two charges associated
with the solution can be written as, \be\label{ZblackD1D5} Z =
\rho_0 \cosh^{\frac{1}{2}}\theta_1 \cosh^{\frac{1}{2}}\theta_5,
\qquad Q_1 = \rho_0^2 \sinh2\theta_1, \qquad Q_5 = \rho_0^2
\sinh2\theta_5. \ee The corresponding KK BON solution is given in
\eqn{bubble}. We have found that to avoid conical singularity at
$\rho_0$, $x^5$ must be periodic with a periodicity given in
\eqn{period}. As before we denote the various bubble quantities with
a subscript `$b$' and also, the bubble parameters with a `tilde'.
So, the periodicity, the size of the bubble, and the fluxes can be
written as follows: \bea\label{LZD1D5BON} L_b &=& 2\pi \rh_0 \cosh
\thet_1 \cosh \thet_5, \quad Z_b = \rh_0 \cosh^{\frac{1}{2}} \thet_1
\cosh^{\frac{1}{2}} \thet_5, \quad Q_{1b} = \rh_0^2 \sinh
2\thet_1,\nn Q_{5b} &=& \rh_0^2 \sinh 2\thet_5. \eea Since $L\gg
\bar{l}_s$, we get from \eqn{LblackD1D5} \be\label{thetone} \cosh
\theta_1 \cosh \theta_5 \gg 1. \ee We have discussed all the
possible cases in a bit detail in the previous subsection and we
could do that in this subsection as well, but instead we will
consider only the case where all the angles are large. This is not
completely unreasonable since given $Q_1$ and $Q_5$, we can always
insist large $L$ to make all the angles $\theta_{1,5}$ and
$\thet_{1,5}$ large. So, for simplicity in this subsection we will
consider only the case where all angles are large.

Now since the angles are large we have from \eqn{LblackD1D5} and
\eqn{ZblackD1D5}, \be\label{blackD1D5condition} e^{\theta_1 +
\theta_5} = \left(\frac{2L}{\bar{l}_s}\right)^2, \quad \frac{L}{Z} =
\frac{\bar{l}_s}{\rho_0}, \quad Q_1=\frac{\rho_0^2}{2}
e^{2\theta_1}, \quad Q_5 = \frac{\rho_0^2}{2} e^{2\theta_5} \ee and
from \eqn{LZD1D5BON}, \be\label{BOND1D5condition} L_b = \frac{\pi
\rh_0}{2} e^{\thet_1 + \thet_5}, \quad \frac{L_b}{Z_b} = \pi
e^{\frac{\thet_1 + \thet_5}{2}}, \quad Q_{1b} = \frac{\rh_0^2}{2}
e^{2\thet_1}, \quad Q_{5b} = \frac{\rh_0^2}{2} e^{2\thet_5}. \ee Now
for closed string tachyon condensation occurring on the horizon, as
mentioned earlier, we must equate, \be\label{condition3} L=L_b,
\qquad Z \approx Z_b, \qquad Q_1 = Q_{1b}, \qquad {\rm and} \qquad
Q_5 = Q_{5b}. \ee Using the relation $L/Z \approx L_b/Z_b$, we have
\be\label{rhozero} \rho_0 = \frac{\bar{l}_s}{\pi} e^{-\frac{\thet_1
+ \thet_5}{2}} \ll \bar{l}_s \ee Further from the expression of $Z$
given in \eqn{ZblackD1D5} and the requirement of $Z \gg \bar l_s$
for small curvature at the horizon, we have, using \eqn{rhozero},
\be\label{Z} Z = \frac{\bar{l}_s}{2\pi}
\left(\frac{e^{\theta_1+\theta_5}}{e^{\thet_1 +
    \thet_5}}\right)^{\frac{1}{2}} \gg \bar{l}_s.
\ee Eq.\eqn{Z} implies $e^{\theta_1 + \theta_5} \gg e^{\thet_1 +
\thet_5}$. Also from $Q_1/Q_5 = Q_{1b}/Q_{5b}$, we obtain
$e^{\theta_1 - \theta_5} = e^{\thet_1
  - \thet_5}$ and combining these two relations we have
\be\label{condition4} e^{\theta_1} \gg e^{\thet_1}, \qquad
e^{\theta_5} \gg e^{\thet_5}. \ee Using \eqn{condition4}  and the
charge conservation we get \be\label{condition5} \rho_0 \ll \rh_0.
\ee We can now solve $\theta_1$, $\theta_5$ and $\rho_0$ from
\eqn{blackD1D5condition} as, \be\label{solution1} e^{\theta_1} =
\frac{2L}{\bar{l}_s}\left(\frac{Q_1}{Q_5}\right)^{\frac
{1}{4}},\quad e^{\theta_5} =
\frac{2L}{\bar{l}_s}\left(\frac{Q_5}{Q_1}\right)^{\frac{1}{4}},
\quad \rho_0 = \frac{\bar{l}_s}{\sqrt{2}} \left(\frac{Q_1
    Q_5}{L^4}\right)^{\frac{1}{4}}
\ee and solve $\thet_1$, $\thet_5$ and $\rh_0$ from
\eqn{BOND1D5condition} as, \be\label{solution2} e^{\thet_1} =
\frac{\sqrt{2}}{\pi} \sqrt{\frac{L^2}{Q_5}}, \quad e^{\thet_5} =
\frac{\sqrt{2}}{\pi} \sqrt{\frac{L^2}{Q_1}}, \quad \rh_0 = \pi L
\left(\frac{Q_1 Q_5}{L^4}\right)^{\frac{1}{2}}. \ee It can now be
easily checked that with the above solutions the conditions for the
closed string tachyon condensation \eqn{condition3},
\eqn{condition4} and \eqn{condition5} can be satisfied provided,
\be\label{provisional} (Q_1 Q_5)^{\frac{1}{4}}  \gg \frac{\bar
l_s}{\sqrt{2}\, \pi}. \ee Let us try to understand the relation
\eqn{provisional} in more detail here. Writing $Q_1 = {\bar g}_s
(N_1/{\bar v}_4){\bar\alpha}'$ and $Q_5 = {\bar g}_s N_5 {\bar
\alpha}'$, where ${\bar g}_s$ is the asymptotic string coupling and
${\bar \alpha}' = \bar {l}_s^2$, with $\bar {l}_s$ the fudamental
string length. Also $\bar {v}_4 = V_4/\bar{\alpha}'^2$. As before we
have put a `bar' in string coupling to distinguish from that used in
the previous subsection and they are related by S-duality. Note that
the total volume $V_4$ remains unchanged. Now using these $Q_1$ and
$Q_5$, the relation \eqn{provisional} becomes,
\be\label{intermediate} \frac{N_1 \bar{g}_s^2}{\bar{v}_4} \gg
\frac{1}{4\pi^4 N_5}. \ee

Let us now look at the dilaton which must be small at $\rho=\rho_0$.
The dilaton has the form (see eq.\eqn{bubble}) $e^{2\phi} =
\bar{g}_s^2 \G_1/\G_5 = \bar{g}_s^2 \cosh^2\theta_1/\cosh^2\theta_5
\approx \bar{g}_s^2 Q_1/Q_5$ for large $\theta$'s. Using the form of
$Q_1$ and $Q_5$ given above we get \be\label{dilaton} e^{2\phi} =
\frac{\bar{g}_s^2 N_1}{\bar{v}_4 N_5} \ll 1 \quad \Rightarrow \quad
\bar{g}_s^2 N_1 \ll \bar{v}_4 N_5. \ee Now using S-duality
$\bar{g}_s = 1/g_s$, $\bar{\alpha}' = g_s \alpha'$ and $V_4 = v_4
\alpha'^2 = \bar{v}_4 \bar{\alpha}'^2 = \bar{v}_4 g_s^2 \alpha'^2
\Rightarrow \bar{v}_4  g_s^2 = v_4$, we get from \eqn{dilaton}
\be\label{smalldilaton} N_5 \gg \frac{N_1}{v_4}. \ee This is the
precise condition for using the S-dual D1/D5 description as
mentioned in the previous subsection.  This relation is also
consistent with the condition of closed string tachyon condensation
we obtained in \eqn{intermediate}. In fact combining
\eqn{intermediate} and \eqn{smalldilaton} we have \be N_5 \gg
\frac{N_1}{v_4} \gg \frac{1}{4\pi^4 N_5}. \ee So, everything fits
nicely.

\section{Special cases}

In this section we will mention two special cases of our general
non-susy F/NS5 solution and non-susy D1/D5 solution for the case of
$k = 1$ as discussed in the previous two sections. First we will
show how the two-charge F-string discussed by Horowitz can be
obtained as a special case of the non-susy F/NS5 solution and then
we show how the interpolating solution between AdS$_3$ black hole
and the global AdS$_3$ can be obtained as a special case of the
non-susy D1/D5 solution.

\subsection{Two-charge F-string}

Two charge F-string solution considered by Horowitz can be seen to
arise as a special case from the general non-susy F/NS5 solution we
obtained in \eqn{twochargeFSch}. The two-charge F-string solution is
a six-dimensional string solution which can be obtained if we simply
restrict the parameters $\delta_1$ and $\delta_2$ as
$\delta_1=4\delta_2$ in \eqn{twochargeFSch} and compactify the
directions $x^1, \ldots, x^4$ on T$^4$. The solution then reduces to
\bea\label{horowitzblack} ds_{\rm 6,str}^2 &=& G_1^{-1}
f^{\frac{{\hat \alpha}_1}{2}} \left(-dt^2 + f^{3\delta_2}
(dx^5)^2\right) + G_5 f^{-\frac{{\hat \alpha}_5}{2} +
\frac{3\delta_2}{2} + \frac{1}{2}} \left(\frac{d\rho^2}{f} + \rho^2
d\Omega_3^2\right),\nn e^{2\tilde \phi} &=& G_5 G_1^{-1}
 f^{-\frac{{\hat \alpha}_5}{2} + \frac{{\hat \alpha}_1}{2} + 3\delta_2},\nn
B_{[2]} &=& \frac{1}{2}\sinh2\theta_1 \left(\frac{1 - f^{\frac
{{\hat \alpha}_{1} + {\hat \beta}_{1}}{2}}} {G_1}\right) dt \wedge
dx^5, \qquad H_{[3]} = b {\rm Vol}(\Omega_3), \eea where the
parameter relations following from \eqn{parametersnew} are given as
${\hat \alpha}_1 - {\hat \beta}_1 = 6\delta_2$, ${\hat \alpha}_5 -
{\hat \beta}_5 =-6\delta_2$ and $({\hat \alpha}_1 + {\hat
\beta}_1)^2 + ({\hat \alpha}_5 + {\hat \beta}_5)^2 = 12
(1-3\delta_2^2)$. The solution \eqn{horowitzblack} is the $D=6$
non-susy two charge F-string solution and is characterized by five
independent parameters $({\hat \alpha}_1 + {\hat \beta}_1)$, $({\hat
\alpha}_5 + {\hat \beta}_5)$, $\rho_0$, $\theta_1$ and $\delta_2$.
In the above we have put $\theta_1=\theta_5$ as Horowitz for
simplicity. It can be easily checked that the solution
\eqn{horowitzblack} interpolates between two-charge black F-string
and KK BON if we vary the parameters keeping $\rho_0$ and $\theta_1$
fixed. Indeed if we choose \be\label{horblackparameters} {\hat
\alpha}_1 + {\hat \beta}_1 = 2, \qquad {\hat \alpha}_5 + {\hat
\beta}_5 =2, \qquad \delta_2 = -\frac{1}{3} \ee (which implies from
the parameter relations given above that ${\hat \alpha}_1 = 2$,
${\hat \beta}_1 = 0$ and ${\hat \alpha}_5 = 0$, ${\hat \beta}_5 =
2$), then the configuration \eqn{horowitzblack} takes the form,
\bea\label{horowitzblack1} ds_{\rm 6,str}^2 &=& \G_1^{-1} \left(-f
dt^2 + (dx^5)^2\right) + \G_1 \left(\frac{d\rho^2}{f} + \rho^2
d\Omega_3^2\right),\nn e^{2\tilde \phi} &=& 1,\nn B_{[2]} &=& \coth
\theta_1 \left(\frac{\G_1-1}{\G_1}\right) dt \wedge dx^5, \qquad
H_{[3]} = b {\rm Vol}(\Omega_3), \eea where $b=\rho_0^2 \sinh
\theta_1$ and $G_1$ and $G_5$ are now equal and take the form
$G_{1,5} \to \G_1 (=\G_5) = 1 + \rho_0^2 \sinh^2 \theta_1/\rho^2$.
\eqn{horowitzblack1} is precisely the two charge black F-string
described in \cite{Horowitz:2005vp}. On the other hand, if we
choose, \be\label{horBONparameters} {\hat \alpha}_1 + {\hat \beta}_1
= 2, \qquad {\hat \alpha}_5 + {\hat \beta}_5 =2, \qquad \delta_2 =
\frac{1}{3} \ee (which implies from the parameter relations given
above that ${\hat \alpha}_1 = 0$, ${\hat \beta}_1 = 2$ and ${\hat
\alpha}_5 = 2$, ${\hat \beta}_5 = 0$), then the configuration
\eqn{horowitzblack} takes the form, \bea\label{horowitzBON} ds_{\rm
6,str}^2 &=& \G_1^{-1} \left(- dt^2 + f (dx^5)^2\right) + \G_1
\left(\frac{d\rho^2}{f} + \rho^2 d\Omega_3^2\right),\nn e^{2\tilde
\phi} &=& 1,\nn B_{[2]} &=& \coth \theta_1
\left(\frac{\G_1-1}{\G_1}\right) dt \wedge dx^5, \qquad H_{[3]} = b
{\rm Vol}(\Omega_3). \eea This is precisely the KK BON solution
given in \cite{Horowitz:2005vp}. Here in order to avoid the conical
singularity at $\rho = \rho_0$, the coordinate $x^5$ must be
periodic with period $L=2\pi\rho_0\cosh^2 \theta_1$. It is therefore
clear that \eqn{horowitzblack} is the solution which interpolates
between the two charge black F-string to KK BON by continuously
varying the parameters ${\hat \alpha}_{1,5}$, ${\hat \beta}_{1,5}$
and $\delta_2$ and there is no need to take the double Wick
rotation. As we have seen for the general case of non-susy F/NS5
solution in sections 3 and 4, that the bubble here could be stable
and static and the interpolation can be understood as a physical
process of closed string tachyon condensation.

\subsection{AdS$_3$ black hole}

The interpolation from the AdS$_3$ black hole to global AdS$_3$ can
be seen to arise as a special case of non-susy D1/D5 system we
obtained in \eqn{D1D5D0Sch}. The general intersecting non-susy D1/D5
system with chargeless D0-branes in Schwarzschild-like coordinate is
given in \eqn{D1D5D0Sch}. If we fix ${\hat \alpha}_{1,5} + {\hat
\beta}_{1,5} = 2$ keeping $\delta_1$ and $\delta_2$ arbitrary (so,
individually ${\hat \alpha}_1$, ${\hat \alpha}_2$, or ${\hat
\beta}_1$, ${\hat \beta}_2$ remain arbitrary) then the functions
$G_{1,5}$ will always have the forms $\G_{1,5} = \cosh^2\theta_{1,5}
- f \sinh^2\theta_{1,5} = 1 + \rho_0^2 \sinh^2\theta_{1,5}/\rho^2$.
Note that this restriction of ${\hat \alpha}_{1,5} + {\hat
\beta}_{1,5}$ is necessary to have the AdS structure unlike in the
previous case. Let us also put $\theta_1=\theta_5$ for simplicity as
in the previous case. Now if we restrict the radial variable in the
region $\rho_0 \le \rho \ll \rho_0 \sinh\theta_1$, then we have
$\G_1 = \G_5 \approx \rho_0^2 \sinh^2\theta_1/\rho^2 \equiv
R^2/\rho^2$. With these restrictions the solution \eqn{D1D5D0Sch}
reduces to, \bea\label{AdS3Sch} ds_{\rm str}^2 &=&
\frac{\rho^2}{R^2}\left(- f^{\frac{1}{2} - \frac{3\delta_1}{8}} dt^2
 + f^{\frac{1}{2} + \frac{\delta_1}{8} + \delta_2}
(dx^5)^2\right) + \frac{R^2}{\rho^2} f^{- \frac{\delta_1}{8} +
\frac{\delta_2}{2}} \left(\frac{d\rho^2}{f} + \rho^2
d\Omega_3^2\right) + \sum_{i=1}^4 (dx^i)^2, \nn e^{2\phi} &=&
f^{-\frac{\delta_1}{4} + \delta_2},\nn F_{[3]} &=&
-2\frac{\rho}{R^2}\,\coth\theta_1\, d\rho \wedge dt \wedge dx^5,\nn
 F_{[7]} &= & -2 \frac{\rho}{R^2}\,\coth\theta_1\, d\rho \wedge dt
\wedge dx^1 \wedge dx^2 \wedge dx^3 \wedge dx^4 \wedge dx^5. \eea
The parameter relations \eqn{parametersnew} in this case take the
forms, \bea\label{AdS3parameters} & & {\hat \alpha}_1 = 1 -
\frac{3}{4}\delta_1, \quad {\hat \alpha}_5 = 1 +
\frac{3}{4}\delta_1, \quad {\hat \beta}_1 = 1 + \frac{3}{4}\delta_1,
\quad {\hat \beta}_5 = 1 - \frac{3}{4}\delta_1,\nn & & \frac{3}{8}
\delta_1^2 + 3\delta_2^2 - 1 = 0. \eea Now it can be checked that if
we choose \be\label{AdS3blackparameter} \delta_1 = -\frac{4}{3}
\qquad {\rm and} \quad \delta_2 = -\frac{1}{3} \ee implying from
\eqn{AdS3parameters} ${\hat \alpha}_1=2$, ${\hat
  \beta}_1=0$, ${\hat \alpha}_5 = 0$, ${\hat \beta}_5 =2$, then the above
solution \eqn{AdS3Sch} reduces to
\bea\label{AdS3blackhole}
ds_{\rm str}^2 &=& \frac{\rho^2}{R^2}\left(-
f dt^2
 +
(dx^5)^2\right) + \frac{R^2}{\rho^2} \left(\frac{d\rho^2}{f} +
\rho^2 d\Omega_3^2\right) + \sum_{i=1}^4 (dx^i)^2, \nn e^{2\phi} &=&
1,\nn F_{[3]} &=& -2\frac{\rho}{R^2} \coth\theta_1 \, d\rho \wedge
dt \wedge dx^5, \nn F_{[7]} &=& -2 \frac{\rho}{R^2}\coth\theta_1\,
d\rho \wedge dt \wedge dx^1 \wedge dx^2 \wedge dx^3 \wedge dx^4
\wedge dx^5. \eea This is the AdS$_3$ $\times$ S$^3$ $\times$ T$^4$
charged black hole solution. On the other hand, if we choose
\be\label{AdS3globalparameter} \delta_1 = \frac{4}{3}, \qquad {\rm
and} \quad \delta_2 = \frac{1}{3} \ee implying from
\eqn{AdS3parameters} ${\hat \alpha}_1=0$, ${\hat
  \beta}_1=2$, ${\hat \alpha}_5 = 2$, ${\hat \beta}_5 =0$, then the above
solution \eqn{AdS3Sch} reduces to \bea\label{globalAdS3} ds_{\rm
str}^2 &=& \frac{\rho^2}{R^2}\left(- dt^2 + f (dx^5)^2\right) +
\frac{R^2}{\rho^2} \left(\frac{d\rho^2}{f} + \rho^2
d\Omega_3^2\right) + \sum_{i=1}^4 (dx^i)^2, \nn e^{2\phi} &=& 1,\nn
F_{[3]} &=& -2\frac{\rho}{R^2} \coth\theta_1\,d\rho \wedge dt \wedge
dx^5,\nn F_{[7]} &=&-2 \frac{\rho}{R^2}\coth\theta_1\, d\rho \wedge
dt \wedge dx^1 \wedge dx^2 \wedge dx^3 \wedge dx^4 \wedge dx^5. \eea
In this case, the conical singularity at $\rho=\rho_0$ can be
avoided if the coordinate $x^5$ has the periodicity
\be\label{periodads} L = 2 \pi \rho_0 \sinh^2\theta_1 \ee This is
the global AdS$_3$ $\times$ S$^3$ $\times$ T$^4$ solution. We thus
find that the classical solution \eqn{AdS3Sch} interpolates nicely
between the AdS$_3$ black hole and the global AdS$_3$ by varying
some parameters characterizing the solution. Again as we have seen
for the general non-susy D1/D5 solution in sections 3 and 4, the
bubble here could be stable and static and the transition can be
understood to be caused by the closed string tachyon condensation.

\section{Discussion}

It has been argued by Horowitz \cite{Horowitz:2005vp} that
black strings under certain conditions can decay into KK BON
by a perturbative stringy process called closed string tachyon condensation.
The two ends of this
process (black string and KK BON) are well described by the classical
supergravity configuration as is well-known. In this paper we have constructed
the more general supergravity solution describing non-susy D1/D5 and non-susy
F/NS5 solutions. These solutions were shown to interpolate smoothly between
black or non-extremal solution and the KK BON by varying some parameters
characterizing the solutions from one set of values to another. Horowitz's
two charge black F-string and AdS$_3$ black hole were shown to arise as
special cases of these general solutions. We have performed a time symmetric
general bubble initial data analysis to argue that the final bubble
configurations could, under certain conditions, be locally stable and static
such that they do not evolve further perturbatively. We have further shown
that this transition can be physically interpreted, under certain
circumstances, as the perturbative stringy process of closed string
tachyon condensation and the interpolating solutions in turn could be
thought of as models of such process.

However, there could be problems with these
interpretations to which we turn next.
\begin{itemize}

\item The interpolating solutions \eqn{twochargeFSch}, \eqn{D1D5D0Sch},
are well-behaved only at the two end points, the black solutions
have singularities masked by regular horizons and the bubble solutions are
completely regular. However, the solutions have naked singularities
at $\rho = \rho_0$ for all the
intermediate points. If they represent the closed string tachyon condensation
how do we interpret the intermediate stages which are singular?

\item It is known \cite{Adams:2005rb}
that the closed string tachyon condensation on the horizon is
a quick process which occurs at the time scale of the order of string scale.
But the interpolation by a classical supergravity description is a continuous
process which is a slow adiabatic process. How can a violent process of closed
string tachyon condensation be described by supergravity?
\end{itemize}

These points have been discussed for black D$p$-branes in
\cite{Lu:2007bu}. We will briefly mention them here. For the first
point, we remark that the intermediate solutions with parameters
other than those given in \eqn{blackFNS5parameter1},
\eqn{blackFNS5parameter2}, \eqn{FNS5bubbleparameter},
\eqn{blackD1D5parameter}, and \eqn{bubbleparameter}, corresponding
to the two end points of various solutions, are all regular in the
region $\rho_0 < \rho \leq \infty$ and the naked singularity at
$\rho=\rho_0$ reflects our inability to describe the system
classically where the violent quantum process of closed string
tachyon condensation is occurring. It is very likely that quantum
mechanically there are no singularities, but classically we do not
have a good description in general for $\rho\leq\rho_0$. To an
observer far away from the core region, only the long range force
would appear and so, we have a classical description of the dynamics
there. The long distance description is just the family of
intermediate solutions with naked singularities. The singularities
are actually the artifact as their appearance is due to the
extrapolation of the solutions valid only at long distance to the
region where the description is invalid and where perhaps we have
quantum process without any singularity.

For the second point we remark that it is true that the time scale for the
completion of the closed string tachyon condensation process at the horizon
is of the order of string scale, but this time is the local proper time. Due
to the red-shift factor in front of the time coordinate it is clear that
for an observer at infinity (with respect to whom the ADM mass is measured)
the time taken for the completion of this process would be infinite and so
for this observer the closed string tachyon condensation would appear as a
slow, adiabatic process which can be well described by a supergravity
with smooth interpolation. This is what our interpolating solutions describe
and have no conflict with the observations made in \cite{Adams:2005rb}.

\section*{Acknowledgements:}

 JXL, ZLW and
RJW acknowledge support by grants from the Chinese Academy of
Sciences, a grant from 973 Program with grant No: 2007CB815401 and
grants from the NSF of China with Grant No:10588503 and 10535060.

\end{document}